\documentclass[12pt]{article}
\usepackage[dvips]{color}
\usepackage{amsmath}
\usepackage{graphicx}
\usepackage{subfigure}
\usepackage{epsfig}
\usepackage{amsmath}
\usepackage{cite}
\usepackage{color}
\usepackage{subfigure}
\usepackage{multirow}

\begin{document}
\begin{center}
\Large{\bf  Bulk-boundary and RPS Thermodynamics \\from Topology perspective}\\
 \small \vspace{1cm}
 {\bf Jafar Sadeghi $^{\star}$\footnote {Email:~~~pouriya@ipm.ir}}, \quad
 {\bf Mohammad Reza Alipour $^{\star}$\footnote {Email:~~~mr.alipour@stu.umz.ac.ir}}\quad
 {\bf Saeed Noori Gashti$^{\dag,\star}$\footnote {Email:~~~saeed.noorigashti@stu.umz.ac.ir}}, \quad  \small \vspace{0.3cm}
 \hspace{2cm} {\bf Mohammad Ali S. Afshar $^{\star}$\footnote {Email:~~~m.a.s.afshar@gmail.com}}\quad\\
\vspace{0.5cm}$^{\star}${Department of Physics, Faculty of Basic
Sciences,\\
University of Mazandaran
P. O. Box 47416-95447, Babolsar, Iran}\\
\vspace{0.5cm}$^{\dag}${School of Physics, Damghan University, P. O. Box 3671641167, Damghan, Iran}
\small \vspace{1cm}
\end{center}
\begin{abstract}
In this article, we investigate the bulk-boundary and restricted phase space (RPS) thermodynamics of Rissner-Nordström (R-N) AdS and 6-dimensional charged Gauss-Bonnet AdS black holes. Additionally, we examine the topological characteristics of the considered black holes and compare them with the results of extended thermodynamics. We have found that the topological behavior of the bulk-boundary thermodynamics is the same as that of the extended thermodynamics, whereas the RPS thermodynamics exhibits a distinct behavior. Furthermore, we demonstrate that within the RPS formalism, there is only one critical point with a topological charge of +1 $(Q_t=+1)$. Moreover, in the RPS formalism, the inclusion of higher derivative curvature terms in the form of Gauss-Bonnet gravity does not alter the topological classification of critical points in charged AdS black holes.
\end{abstract}

\tableofcontents
\section{Introduction}
Thermodynamics of black holes is the study of how black holes behave as thermodynamic systems, including their temperature, entropy, heat capacity, and other properties. One of the main challenges is to understand the phase transitions and critical points of black holes in different contexts, such as anti-de Sitter (AdS) or de Sitter (dS) space, with or without electric charge, nonlinear electromagnetic fields, etc. Some important concepts in thermodynamics are critical points and phase transitions. A phase transition is a change of state of matter from one phase to another, such as from liquid to gas or from solid to liquid \cite{a,b,c,d,e,f,g,h,i,j,k,m,n,o,p}. In thermodynamics, phase transitions are characterized by discontinuities or singularities in some thermodynamic quantities, such as pressure, volume, temperature, etc. Phase transitions can be classified into different orders depending on the behavior of these quantities near the transition point. A recent approach to studying the thermodynamics and phase transitions of black holes is to use Duan's topological current-mapping theory. This theory introduces topology to the study of black hole thermodynamics by assigning a topological charge to each critical point in the phase diagram. This method can reveal new features and classifications of black hole thermodynamics that are not captured by conventional methods. For example, it can distinguish between conventional and novel critical points, which have different implications for the first-order phase transition. It can also classify different types of black holes into different topological classes based on their topological charges. Many important works have been done so far on investigating thermodynamics from the point of view of topology, exploring different aspects of this topic. For further study and research, you can refer to the following references: \cite{aa,bb,cc,dd,ee,ff,ff1,ff2,ff3,ff4,ff5,gg,hh,ii,jj,kk,kk1,ll,1000,1001,1002,mm,mm1,nn,oo,pp,pp1,pp2,pp3}. In this article, we aim to conduct thermodynamic studies by comparing two different spaces, namely Bulk-Boundary and Restricted Phase Space, for different types of black holes, which have not been investigated so far. We will compare the results with each other and with other works in the literature.\\\\ 
Restricted phase space thermodynamics (RPST) is a new formalism for studying the thermodynamics of AdS black holes. The construction is based on Visser's holographic thermodynamics, but with the AdS radius being fixed. This means that the pressure and the conjugate volume are excluded from the thermodynamic variables, and only the central charge $C$ of the dual CFT and the chemical potential $\mu$ are included.
The RPST formalism has several advantages over the conventional Extended Phase Space (EPS) formalism, which treats the cosmological constant as pressure and introduces an extra pair of state variables (P, V). For example, the RPST formalism can avoid some ambiguities and inconsistencies in the EPS formalism, such as the definition of mass, the Smarr relation, and the first law of thermodynamics. The RPST formalism can also automatically satisfy the Euler relation, and it explicitly demonstrates the first-order homogeneity of mass and the zeroth-order homogeneity of the intensive variables.
The RPST formalism can be applied to various types of AdS black holes, such as Kerr-AdS and RN-AdS \cite{fff,ggg}, and reveal some interesting thermodynamic behaviors and phase transitions. For example, it can show that there is a first-order supercritical phase equilibrium in the $T-S$ processes at a fixed nonvanishing charge or angular momentum, while at vanishing charge or angular momentum or at fixed potentials, there is always a non-equilibrium transition from a small unstable black hole state to a large stable black hole state. Moreover, there is a Hawking-Page phase transition in the $\mu-C$ processes. The RPST formalism can also classify different types of black holes into different topological classes based on their topological charges of critical points \cite{h,bbb,ccc,ddd,eee,fff,ggg}.\\\\
Bulk-boundary thermodynamic equivalence is a concept that states that the thermodynamics and phase transitions of AdS black holes in the bulk gravity theory are equivalent to those of the dual CFT on the boundary, from the point of view of topology. This means that the bulk and boundary thermodynamics have the same topological charges of critical points, which can be calculated using the residue method.
This concept was proposed by Zhang and Jiang in their paper \cite{100} "Bulk-boundary thermodynamic equivalence: a topology viewpoint". In their work, they set both the cosmological constant and Newton's constant to be dynamical, and developed mass/energy formulas in terms of thermodynamic state functions for three different scenarios: the extended thermodynamics, which includes both pressure and volume; the mixed thermodynamics, which includes only pressure; and the boundary CFT thermodynamics, which includes neither pressure nor volume. They applied these formulas to study the thermodynamics and phase transitions of charged AdS black holes in various dimensions and with various charges or angular momenta, and compared them with other works in the literature.
They used the residue method to study the topological properties of phase transitions. They defined a vector field $\varphi$ on a complex plane $z$, which is related to the temperature T and entropy S. They also defined a complex function $\Omega$ on $z$, which is related to different thermodynamic potentials for extended thermodynamics, mixed thermodynamics, and boundary CFT thermodynamics. They used these definitions to calculate the residues of $\varphi/\Omega$ at different critical points on the $z$-plane and used them as topological charges to classify different types of phase transitions.
They found that the bulk and boundary thermodynamics are topologically equivalent for both criticalities and first-order phase transitions in the canonical ensembles, as well as for the Hawking-Page (like) phase transitions in the grand canonical ensembles. They also found that these three kinds of phase transitions are distinguished by their unique topological charges. They claimed that their results exemplify the gravity-gauge duality in terms of topology \cite{100,101,102}.\\\\
Therefore, we will choose different black holes and study their topological behavior in two spaces: bulk-boundary and restricted phase space, and then compare the results.\\
\emph{In \cite{100}, it was shown that the total topological charge remains constant in extended, mixed, and CFT thermodynamics. On the other hand, we show that in RPS thermodynamics, the total topological charge changes in comparison with other thermodynamics. This effect may arise from the fixed cosmological constant.}\\
One of the proposed methods to investigate the thermodynamics of a black hole is the use of topology, which leads to our prognostic of the phase structure of a black hole. Therefore, in this study, we tried to choose black holes so that they have a different phase structure. Since the charged black hole and the (4 and 5)-dimensional Gauss-Bonnet AdS black holes have similar behavior in the phase structure. We choose the charged black hole to investigate the thermodynamic behavior of its topology with the 6-dimensional Gauss-Bonnet AdS black hole in two extended and restricted phase spaces.

Based on the above concepts, we can organize the article as follows.\\
In section 2, we will briefly explain thermodynamic topology. In section 3, we will perform calculations related to thermodynamic topology from the perspective of bulk-boundary for two black holes, namely the R-N AdS black hole and the charged Gauss-Bonnet AdS black hole. In section 4, we will undergo a similar process for restricted Phase Space and compare the results with each other and with the latest findings in the literature. Finally, we will summarize the results in section 5.

\section{Thermodynamic topology}
The critical points of the phase diagram of thermodynamic systems can be classified into three categories: conventional, novel, and neutral, based on the topological charge given by the winding number of Duan's $\phi$-mapping theory. This theory uses a scalar thermodynamic function $\Phi$ and a vector field $\phi$ to study the topological features and stability conditions of black hole solutions in various gravity theories. The zero points of $\phi$ correspond to the critical points of $\Phi$, and the topological charge is determined by the deflection angle of $\phi$ along a closed contour around the zero point. The conventional class has a negative charge, the novel class has a positive charge, and the neutral class has a zero charge. Different gravity theories and parameters may affect the topological class of black hole solutions \cite{aa,200,ff}.
Researchers have recently discovered that black holes can be classified into two types based on their thermodynamic behavior. These types are called conventional and novel, and they are determined by the topological charge of the critical points on the phase diagram. The phase diagram shows how the temperature of the black hole depends on other properties like entropy, pressure, and some extra parameters ($x^{i}$) that vary depending on the gravity theory and the black hole solution. These properties are considered as variables in the extended thermodynamic systems. The temperature function for these systems is given by \cite{aa},
\begin{equation}\label{1}
T=T(S, P, x^{i}) \qquad or \qquad T=T(r_h, P, x^{i}).
\end{equation}
An important fact is that the critical points on the phase diagram of a thermodynamic system, i.e. where the system changes abruptly from one phase to another, can be found by solving the conditions for a stationary point of inflection,
\begin{equation}\label{2}
\bigg(\frac{\partial T}{\partial S}\bigg)_{P,x_i}=0   \qquad or \qquad  \bigg(\frac{\partial T}{\partial r_h}\bigg)_{P,x_i}=0.
\end{equation}
Next, we represent the newly introduced thermodynamic function as $\Phi$, where the pressure $P$ has been removed according to equation \eqref{2},
\begin{equation}\label{3}
\Phi=\frac{1}{\sin\theta}T(S, x^{i}).
\end{equation}
The above equation is obtained by eliminating the variable $P$ using the equation of \eqref{2} and adding the factor $\frac{1}{\sin\theta}$ to simplify the analysis.
They also introduce a new vector field to develop the framework of Duan's $\phi$-mapping theory.
\begin{equation}\label{4}
\phi=(\phi^{S}, \phi^{\theta}),
\end{equation}
where
\begin{equation}\label{5}
\phi^{S}=(\partial_{S}\Phi),\hspace{1cm}\phi^{\theta}=(\partial_{\theta}\Phi).
\end{equation}
Also, the charge can be considered as,
\begin{equation}\label{6}
Q_{t}=\int_{\Sigma}\Sigma_{i=1}^{N}\beta_{i}\eta_{i}\delta^{2}(\overrightarrow{x}-\overrightarrow{z}_{i})d^{2}x=\Sigma_{i=1}^{N}\beta_{i}\eta_{i}=\Sigma_{i=1}^{N}\omega_{i},
\end{equation}
where $\eta_{i}$ is the sign of the zero component of the topological current $J^{0}(\phi/x)_{z_{i}}$ at the i-th zero point of $\phi$, and it can be either $+1$ or $-1$. $\beta_{i}$ and $\omega_{i}$ are positive integers (Hopf index) that count the number of loops and the winding number around the i-th zero point of $\phi$. The equation above shows that the topological charge $Q_{t}$ is non-zero only at the zero points of $\phi$. Thus, each critical point has a topological charge given by the winding number, and it can be positive or negative. This leads to two different topological classes (conventional and novel), where the negative charge corresponds to conventional and the positive charge corresponds to novel. To visualize these topological classes, one can use a plot of $(\theta-r)$ or $(\theta-S)$ with some smooth curves $C$ that go around the zero points in a positive direction.
As a simplification, chooses $C$ to be an ellipse centered at $(r_{0}, \frac{\pi}{2})$ with a parameter $\vartheta$ that ranges from $0$ to $2\pi$ \cite{200},
\begin{equation}\label{7}
r=a\cos\vartheta+r_{0},\hspace{1cm}\theta=b\sin\vartheta+\frac{\pi}{2}.
\end{equation}
Therefore, the topological charge that given by the winding number can be obtained by calculating the deflection $\Omega(\vartheta)$ of the vector field $\phi$. So we will have,
\begin{equation}\label{8}
Q_{t}=\frac{1}{2\pi}\Omega(2\pi),
\end{equation}
where
\begin{equation}\label{9}
\Omega(\vartheta)=\int_{0}^{\vartheta}\epsilon_{ab}n^{a}\partial_{\vartheta}n^{b}d\vartheta.
\end{equation}
Now, we apply the equations mentioned in this section to two types of black holes: R-N AdS black hole and charged Gauss-Bonnet AdS black hole. We use the bulk-boundary and restricted phase space respectively, and describe the results of our work in detail in the following sections.

\section{Bulk-Boundary thermodynamics}
Bulk-boundary thermodynamics is a formalism that relates the thermodynamic quantities of AdS black holes in the bulk to those of the dual CFT on the boundary via the AdS/CFT correspondence. The bulk quantities include the mass, temperature, entropy, angular momentum, and electric potential of the black hole, while the boundary quantities include the energy, temperature, entropy, angular velocity, and chemical potential of the CFT. The bulk-boundary thermodynamics shows that the first law of thermodynamics and the Smarr relation are satisfied by both the bulk and the boundary quantities. The bulk-boundary thermodynamics also reveals some interesting phase transitions and critical behaviors of the AdS black holes and their dual CFT.\\

A notable discovery in the past few years is that a negative cosmological constant $\Lambda$ results in a thermodynamic pressure $P$ that is positive,
\begin{equation}\label{100}
\begin{split}
P=-\frac{\Lambda}{8\pi G}, \qquad     \Lambda=-\frac{(d-1)(d-2)}{2\ell^2}
\end{split}
\end{equation}
where $G$ is the Newton gravitational constant and $\ell$ is the radius of the $d$-dimensional AdS space (setting $\hbar=c=1$).
Building on the AdS/CFT concept and adopting the methodology outlined in \cite{51,3}, we utilize the duality correlation presented in \cite{53} to express the central charge $C$ in the following manner:
\begin{equation}\label{101}
\begin{split}
C=k\frac{\ell^{d-2}}{16\pi G}
\end{split}
\end{equation}
The value of the factor $k$ is influenced by the specific characteristics of the holographic system.
The first law of black hole thermodynamics in the condition that $\ell$ and $G$ are both variables for a black hole with the characteristics of mass $M$, electric charge $Q$, angular momentum $J$, area $A$, and Gauss-Bonnet constant  $\tilde{\alpha}=(d-3)(d-4)\alpha$ is as follows,
\begin{equation}\label{102}
\begin{split}
\delta M=\frac{\kappa}{8\pi G}\delta A+\Omega \delta J-\frac{V}{8\pi G}\delta \Lambda +\Phi \delta Q-\frac{\beta}{G}\delta G +\gamma \delta \tilde{\alpha}
\end{split}
\end{equation}
where $\kappa, \Omega, V, \Phi, \frac{\beta}{G}$ and $\gamma$ are surface gravity, conjugate angular velocity, thermodynamic volume, electric potential, conjugate to $G$, and conjugate variable of $\tilde{\alpha}$, respectively.
When$G$ and $J$ are both variables, according to Refs. \cite{3,51,52}, we can use the definition of $\mathcal{M}=M(A,\sqrt{G}Q,GJ,\Lambda,\tilde{\alpha})$ in bulk, in which case the relation \eqref{102} will be rewritten \cite{3}, 
\begin{equation}\label{103}
\begin{split}
\delta (GM)=\frac{\kappa}{8\pi }\delta A+\Omega \delta(G J)-\frac{V}{8\pi}\delta \Lambda +\sqrt{G}\Phi \delta(\sqrt{G} Q)+\gamma G \delta \tilde{\alpha}
\end{split}
\end{equation}
By dividing each side by the Newton's constant $G$, we can derive the variation in mass M as, 
\begin{equation}\label{104}
\begin{split}
\delta M=\Omega \delta J +\Phi \delta Q+\frac{1}{G}\big(\frac{\kappa}{8\pi }\delta A-\frac{V}{8\pi}\delta \Lambda+\gamma G \delta \tilde{\alpha} \big)+\frac{1}{G}\big(-M+\Omega J+\frac{Q\Phi}{2} \big)\delta G
\end{split}
\end{equation}
By comparing relations \eqref{102} and \eqref{104}, beta can be obtained,
\begin{equation}\label{105}
\begin{split}
\beta=M+\frac{Q\Phi}{2}-\Omega J
\end{split}
\end{equation}
Also, in order for the central charge to play a role in the first law of thermodynamics, we need to use relations \eqref{100} and \eqref{101}, so, we have \cite{51},
\begin{equation}\label{106}
\begin{split}
\frac{\delta G}{G}=-\frac{2}{d}\frac{\delta C}{C}-(1-\frac{2}{d})\frac{\delta P}{P}
\end{split}
\end{equation}
Therefore, by using equations \eqref{102}, \eqref{104}, \eqref{106} , and $T=\frac{\kappa}{2\pi},  S=\frac{A}{4G}$, we can obtain the first law of thermodynamics in the form of bulk and boundary as follows \cite{3,51}:
\begin{equation}\label{107}
\begin{split}
\delta M=T\delta S+\Omega \delta J+\Phi \delta Q+\mu_c \delta C +V_c\delta P+\gamma \delta \tilde{\alpha} ,
\end{split}
\end{equation}
where
\begin{equation}\label{108}
\begin{split}
\mu_c=\frac{2(V_c-V)}{d-2}\frac{P}{C}, \qquad    V_c=\frac{2M+(d-4)\Phi Q+4\gamma\tilde{\alpha}}{2Pd}
\end{split}
\end{equation}
are the chemical potential $\mu$  and new thermodynamic volume $V_c$.
Next, we examine the topology thermodynamics of two R-N AdS and charged Gauss-Bonnet AdS black holes in the bulk-boundary space.

\subsection{Reissner-Nordström AdS black hole}
A R-N AdS black hole is a type of charged black hole that has a negative cosmological constant. It is a solution to the Einstein-Maxwell equations that describe the gravitational field of a charged, non-rotating, spherically symmetric body. It has two event horizons: an inner and an outer one. The R-N AdS black hole has some interesting properties, such as casting a shadow, exhibiting a phase transition similar to the van der Waals system, and having quantum gravitational corrections to its metric. Now we apply the equations mentioned in the section 2 in R-N AdS black hole. The metric of the R-N AdS black hole is given by \cite{fff}:

$$ds^2 = -f(r)dt^2 + \frac{dr^2}{f(r)} + r^2 d\Omega^2_2$$

where $d\Omega^2_2$ is the metric of a unit two-sphere, and

$$f(r) = 1 - \frac{2GM}{r} + \frac{GQ^2}{4r^2} + \frac{r^2}{\ell^2}$$

where $M$ and $Q$ are related to the mass and charge of the black hole respectively, and $\ell$ is the radius of the AdS space.
So, the temperature of this black hole is expressed in the following form
\begin{equation}\label{10}
\begin{split}
T=\frac{-G \ell^2 Q^2+\ell^2  r_h^2+3  r_h^4}{4 \pi  \ell^2  r_h^3}
\end{split}
\end{equation}
We remove the $\ell$ parameter by setting $\frac{\partial T}{\partial r_h}=0$ and consider the $\Phi$ function as follows,
\begin{equation}\label{11}
\begin{split}
\Phi=\frac{r_h^2-2 G Q^2}{2 \pi  r_h^3 \sin (\theta )}
\end{split}
\end{equation}
Then, using relation \eqref{11}, we obtain vector field $\phi^r=\frac{\partial \Phi}{\partial r_h}$ and $\phi^{\theta}=\frac{\partial \Phi}{\partial \theta}$,
\begin{equation}\label{12}
\begin{split}
\phi^r=\frac{\csc (\theta ) \left(6 G Q^2-r_h^2\right)}{2 \pi  r_h^4},  \qquad \qquad   \phi^{\theta}=\frac{\cot (\theta ) \csc (\theta ) \left(2 G Q^2-r_h^2\right)}{2 \pi  r_h^3}
\end{split}
\end{equation}
The normalized vector field is therefore $n=\bigg(\frac{\phi^r}{\|\phi\|}, \frac{\phi^{\theta}}{\|\phi\|}  \bigg)$. So, we have,
\begin{equation}\label{13}
\begin{split}
& n^r=\frac{ \left(6 G Q^2-r_h^2\right)}{\sqrt{r_h^2 \cot ^2(\theta ) \left(r_h^2-2 G Q^2\right)^2+\left(r_h^2-6 G Q^2\right)^2}}\\
& n^{\theta}=\frac{r_h \cot (\theta ) \left(2 G Q^2-r_h^2\right)}{\sqrt{r_h^2 \cot ^2(\theta ) \left(r_h^2-2 G Q^2\right)^2+\left(r_h^2-6 G Q^2\right)^2}}
\end{split}
\end{equation}
To determine the topological charge of a critical point (where $\phi^r=\phi^{\theta}=0$ ), it is necessary to find its winding number $w_i$. For this purpose, we seek assistance from the orthogonal $\theta-r_h$ plane. In the following discussion, we set $P=1$ and $\ell=\sqrt[4]{\frac{3 C}{8 \pi }}$.
\begin{figure}[h!]
 \begin{center}
 \includegraphics[height=6cm,width=7cm]{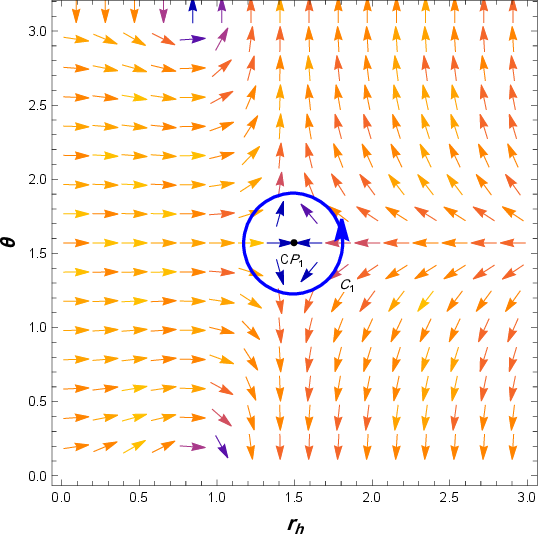}
 \caption{\small{The vector field $n$ is represented by the arrows on the $r_h-\theta$ plane. With the $Q=1, C=1, P=1$}}
 \label{fig1}
 \end{center}
 \end{figure}
As shown in Figure \eqref{fig1}, the topological charge for the R-N AdS black hole in bulk-boundary thermodynamics is $Q_t=-1$. In fact, this finding aligns with the topological charge obtained in the extended black hole thermodynamics outlined in \cite{aa}.

\newpage
\subsection{Charged Gauss-Bonnet AdS black holes}
A charged Gauss-Bonnet AdS black hole is a type of charged black hole that has a negative cosmological constant and a Gauss-Bonnet term in the action. The charged Gauss-Bonnet AdS black hole has some interesting properties, such as having a thermodynamic geometry, exhibiting a phase transition similar to the R-N AdS black hole, and having effects on the shadow, energy emission rate and quasinormal modes of the black hole. The metric of the 6-dimensional charged Gauss-Bonnet AdS black hole is given by \cite{3,bb},

$$ds^2 = -f(r)dt^2 + \frac{dr^2}{f(r)} + r^2(d\theta^2+\sin^2\theta d\phi^2+\cos^2\theta d\Omega^2_2)$$

 and,

$$f(r) = 1 + \frac{r^2}{2\tilde{\alpha}}\left(1 - \sqrt{1 + \frac{16 \pi \tilde{\alpha} G M}{\Sigma_k r^5} - \frac{2\tilde{\alpha} G Q^2}{3r^8} - \frac{4\tilde{\alpha}}{\ell^2}}\right)$$

where $\tilde{\alpha}=6\alpha$ $M$, $Q$ are related to the mass and charge of the black hole respectively, $\alpha$ is the Gauss-Bonnet coupling parameter, and $\ell$ is the radius of the AdS space. To simplify, the area of a 6-dimensional unit sphere is denoted as $\Sigma_k=1$.
We use the articles \cite{3,bb} and have the temperature of the charged Gauss-Bonnet AdS black holes in 6 dimensions,
\begin{equation}\label{14}
\begin{split}
T=\frac{\tilde{\alpha} -\frac{G Q^2}{2 r_h^4}+\frac{5 r_h^4}{\ell^2}+3 r_h^2}{4 \pi  r_h \left(2 \tilde{\alpha} +r_h^2\right)}
\end{split}
\end{equation}
By setting $\frac{\partial T}{\partial r_h}=0$, we eliminate the $\ell$ parameter and focus on the $\Phi$ function,
\begin{equation}\label{15}
\begin{split}
\Phi=\frac{1}{\sin (\theta )}\bigg(\frac{-2 G Q^2+3 r_h^6+2 \tilde{\alpha}  r_h^4}{ 2 \pi  r_h^7+12 \pi  \tilde{\alpha}  r_h^5}\bigg)
\end{split}
\end{equation}

and
\begin{equation}\label{16}
\begin{split}
&\phi^{r_h}=\frac{\partial\Phi}{\partial r_h}=\frac{\csc (\theta ) \left(2 G Q^2 \left(30 \tilde{\alpha} +7 r_h^2\right)-3 r_h^4 \left(r_h^2-2 \tilde{\alpha} \right)^2\right)}{2 \pi  r_h^6 \left(6 \tilde{\alpha} +r_h^2\right)^2},  \\
&\phi^{\theta}=\frac{\partial\Phi}{\partial \theta}=\frac{\cot (\theta ) \csc (\theta ) \left(2 G Q^2-3 r_h^6-2 \tilde{\alpha}  r_h^4\right)}{2 \pi  r_h^7+12 \pi  \tilde{\alpha}  r^5}
\end{split}
\end{equation}
We also obtain the normalized vector field as follows,
\begin{equation}\label{17}
\begin{split}
& n^{r_h}=\frac{2 G Q^2 \left(30 \tilde{\alpha} +7 r_h^2\right)-3 r_h^4 \left(r_h^2-2 \tilde{\alpha} \right)^2}{\sqrt{\left(3 r_h^4 \left(r_h^2-2 \tilde{\alpha} \right)^2-2 G Q^2 \left(30 \tilde{\alpha} +7 r_h^2\right)\right)^2+r_h^2 \cot ^2(\theta ) \left(6 \tilde{\alpha} +r_h^2\right)^2 \left(-2 G Q^2+3 r_h^6+2 \tilde{\alpha}  r_h^4\right)^2}}\\
& n^{\theta}=\frac{r_h \cot (\theta ) \left(6 \tilde{\alpha} +r_h^2\right) \left(2 G Q^2-3 r_h^6-2 \tilde{\alpha}  r_h^4\right)}{\sqrt{\left(3 r_h^4 \left(r_h^2-2 \tilde{\alpha} \right)^2-2 G Q^2 \left(30 \tilde{\alpha} +7 r_h^2\right)\right)^2+r_h^2 \cot ^2(\theta ) \left(6 \tilde{\alpha} +r_h^2\right)^2 \left(-2 G Q^2+3 r_h^6+2 \tilde{\alpha}  r_h^4\right)^2}}
\end{split}
\end{equation}
Then, to determine the topological charge of the critical points, we draw the $r_h-\theta$ diagram using equation \eqref{17}.
\begin{figure}[h!]
 \begin{center}
 \subfigure[]{
 \includegraphics[height=5cm,width=5cm]{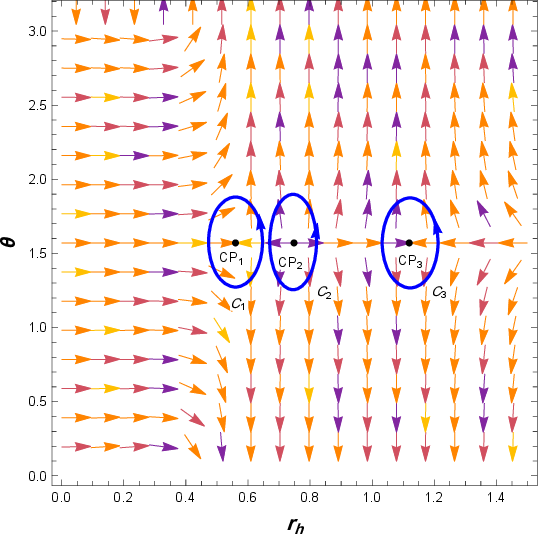}
 \label{fig2a}}
 \subfigure[]{
 \includegraphics[height=5cm,width=5cm]{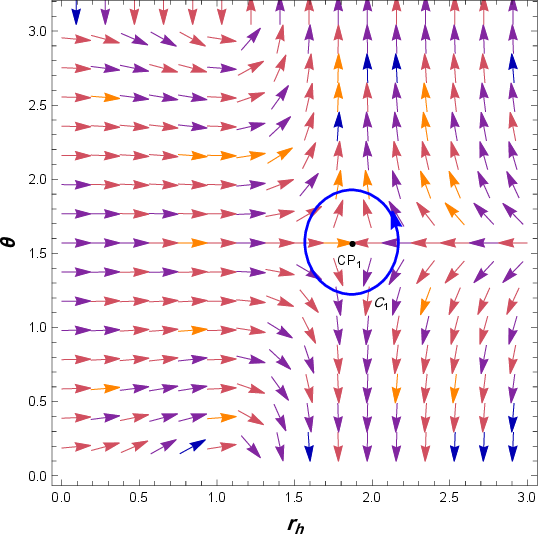}
 \label{fig2b}}
 \caption{\small{The vector field $n$ is represented by the arrows on the $r_h-\theta$ plane. With the $C=30, P=0.015, \tilde{\alpha}=0.5$ and  (a) $Q=0.04$, (b) $Q=1$.}}
 \label{fig2}
 \end{center}
 \end{figure}
As shown in \cite{3}, for the 6D charged Gauss-Bonnet AdS black hole in bulk-boundary  thermodynamics, when $Q<Q_3$, we have three critical points, and when $Q>Q_3$, we have only one critical point.
According to Figure \eqref{fig2}, we find that in all cases in bulk-boundary thermodynamics, the total topological charge is -1 $(Q_t=-1)$.
This result is consistent with \cite{bb} that the total topological charge of the 6D charged Gauss-Bonnet AdS black hole in the extended thermodynamics is -1 $(Q_t=-1)$ in all cases.
\newpage
\section{Restricted Phase Space}
Restricted phase space (RPS) is a new formalism for thermodynamics of AdS black holes that excludes the pressure and the conjugate volume as thermodynamic variables, but includes the central charge of the dual CFT and the chemical potential as intensive variables. The RPS formalism is based on Visser’s holographic thermodynamics, but with the AdS radius must be fixed. The RPS formalism has some advantages over the extended Phase Space (EPS) formalism, such as avoiding the ambiguity of the Smarr relation and the Euler relation, and making explicit the scaling properties of the equations of state. The RPS formalism has been applied to various AdS black hole solutions, such as R-N AdS, Kerr-AdS, and revealed some interesting thermodynamic behaviors, such as phase transitions, criticalities, and Hawking-Page transitions. Now, we will perform the same process as in the previous sections for the mentioned black hole and this time in restricted phase space. We will compare the results with other works in the literature.

Visser has developed a new version of the extended phase space formalism, which takes into account the AdS/CFT correspondence \cite{52,54}. This innovation includes the addition of the CFT central charge $C$ and the conjugate chemical potential $\mu$ as new thermodynamic parameters. The volume and pressure are also changed to that of the CFT, where $\mathcal{V}$ is proportional to $\ell^{d-2}$ (with $\ell$ being the AdS radius) and $\mathcal{P}$ is determined by the CFT equation of states (EOS) 
$E = (d -2)\mathcal{P}\mathcal{V}$, where $d$ represents the dimension of the bulk spacetime.
In RPS thermodynamics, the relationship between the central charge of the black hole and Newton's constant is as follows,
\begin{equation*}\label{20011}
C=\frac{\ell^{d-2}}{G}
\end{equation*}
While both $\ell$ and $G$ are considered to be variable in \cite{52}, the RPS formalism keeps $\ell$ fixed and only $G$ is allowed to vary. Physically, $G$  is proportional to $(\ell_p^{d-2})$, where $\ell_p$ represents the Planck length. Therefore, $C$ has an intuitive interpretation as the number of pieces of the size of Planck “area” that a hypersurface of radius $\ell$ can be divided into, with each such piece representing a single microscopic degree of freedom \cite{57}. Varying $G$ implies changing the Planck length,  which in turn changes the number of pieces mentioned above \cite{57}.
Since we are considering Gauss-Bonnet and R-N AdS black holes, the first law of thermodynamics in this formalism is
\begin{equation}\label{200}
\begin{split}
dE=TdS+\Omega dJ+\mu dC+\hat{\Phi} d\hat{Q} +\gamma d\tilde{\alpha}-\mathcal{P} d\mathcal{V}
\end{split}
\end{equation}
The electric potential and electric charge, $\tilde{Q}$ and $\tilde{\Phi}$ respectively, have been appropriately rescaled.
In the restricted phase space proposed in \cite{fff,ggg}, the radius AdS is considered fixed, in which case the first law of thermodynamics is rewritten as \cite{fff,ggg,56},
\begin{equation}\label{201}
\begin{split}
dM=TdS+\Omega dJ+\mu dC+\hat{\Phi} d\hat{Q} +\gamma d\tilde{\alpha}
\end{split}
\end{equation}
The re-scaled electric potential, $\hat{\Phi}$, and the re-scaled electric charge, $\hat{Q}$, are defined by the dual CFT quantities \cite{fff,ggg,51,56},
\begin{equation}\label{202}
\begin{split}
\hat{Q}=Q\frac{\ell^{\frac{d-2}{2}}}{\sqrt{G}}=Q\sqrt{C}, \qquad  \hat{\Phi}=\Phi \frac{\sqrt{G}}{\ell^{\frac{d-2}{2}}}=\frac{\Phi}{\sqrt{C}}
\end{split}
\end{equation}
In the following, we examine the topology thermodynamics for the considered black holes in the restricted phase space.

\subsection{Reissner-Nordström AdS black hole}
We use \cite{fff} and for temperature, in the RPS formalism, we have
\begin{equation}\label{18}
\begin{split}
T=\frac{\pi  C S-\pi ^2 \hat{Q}^2+3 S^2}{4 \pi ^{3/2} \ell S \sqrt{C S}}
\end{split}
\end{equation}
First, we remove the $\hat{Q}$ parameter when the condition  $\frac{\partial T}{\partial S}=0$ is met and form the $\phi$  function as follows,
\begin{equation}\label{19}
\begin{split}
\Phi=\frac{1}{\sin (\theta )}T(S,C)=\frac{\pi  C+6 S}{\sin (\theta ) \left(6 \pi ^{3/2} \ell \sqrt{C S}\right)}
\end{split}
\end{equation}
Then, utilizing the relation of \eqref{19}, we construct the vector field,
\begin{equation}\label{20}
\begin{split}
&\phi^S=-\frac{C (\pi  C-6 S) \csc (\theta )}{12 \pi ^{3/2} \ell (C S)^{3/2}},  \\
&\phi^{\theta}=-\frac{(\pi  C+6 S) \cot (\theta ) \csc (\theta )}{6 \pi ^{3/2} \ell \sqrt{C S}}
\end{split}
\end{equation}
and
\begin{equation}\label{21}
\begin{split}
&n^S=\frac{\partial \phi^S}{\partial S}=\frac{6 S-\pi  C}{\sqrt{4 S^2 (\pi  C+6 S)^2 \cot ^2(\theta )+(\pi  S-6 S)^2}},  \\
&n^{\theta}=\frac{\partial \phi^{\theta}}{\partial \theta}=-\frac{2 S (\pi  C+6 S) \cot (\theta )}{\sqrt{4 S^2 (\pi  C+6 S)^2 \cot ^2(\theta )+(\pi  C-6 S)^2}}
\end{split}
\end{equation}
According to equation \eqref{21}, we determine that the critical point is $(S, \theta)=(\frac{C}{6},\frac{\pi}{2})$. The obtained result is consistent with \cite{fff}.
\begin{figure}[h!]
 \begin{center}
 \includegraphics[height=6cm,width=7cm]{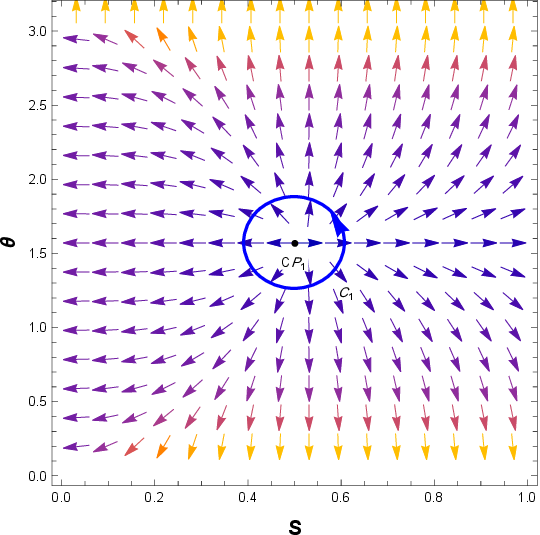}
 \caption{\small{The vector field $n$ is represented by the arrows on the $S-\theta$ plane. With the $C=1.$}}
 \label{fig3}
 \end{center}
 \end{figure}
As shown in Figure \eqref{fig3}, this critical point possesses a topological charge of +1 $Q_t=+1$. This finding contradicts \cite{aa}, implying that in the RPS formalism, it results in a modification of the topological charge.
\newpage
\subsection{Charged Gauss-Bonnet AdS black holes}
We use the article \cite{3} and with the definition $\hat{Q}=Q \sqrt{C}$ \cite{fff,55} and we have,
\begin{equation}\label{22}
\begin{split}
T=\frac{\tilde{\alpha} -\frac{G \hat{Q}^2}{2 C r_h^4}+\frac{5 r_h^4}{\ell^2}+3 r_h^2}{4 \pi  r_h \left(2 \tilde{\alpha} +r_h^2\right)}
\end{split}
\end{equation}
Then, by utilizing relation \eqref{2}, we eliminate the $\hat{Q}$ parameter. Furthermore, in accordance with relation \eqref{2}, we obtain the function,
\begin{equation}\label{23}
\begin{split}
\Phi=\frac{1}{\sin (\theta )}\bigg(\frac{2 \tilde{\alpha}  \ell^2+9 \ell^2 r_h^2+20 r_h^4}{14 \pi  \ell^2 r_h^3+20 \pi  \tilde{\alpha}  \ell^2 r_h}\bigg)
\end{split}
\end{equation}

We also have normalized vector fields,
\begin{equation}\label{24}
\begin{split}
&\phi^{r_h}=-\frac{\csc (\theta ) \left(\ell^2 \left(20 \alpha ^2+63 r_h^4-48 \alpha  r_h^2\right)-20 \left(7 r_h^6+30 \alpha  r_h^4\right)\right)}{2 \pi  l^2 r_h^2 \left(10 \alpha +7 r_h^2\right)^2},  \\
&\phi^{\theta}=-\frac{\cot (\theta ) \csc (\theta ) \left(2 \alpha  \ell^2+9 \ell^2 r_h^2+20 r_h^4\right)}{14 \pi  \ell^2 r_h^3+20 \pi  \alpha  \ell^2 r_h}
\end{split}
\end{equation}

\begin{figure}[h!]
 \begin{center}
 \includegraphics[height=6cm,width=7cm]{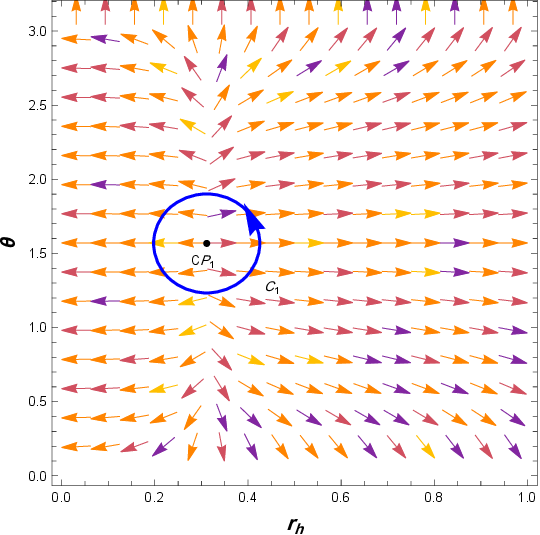}
 \caption{\small{ The vector field $n$ is represented by the arrows on the $r_h-\theta$ plane. With the $\ell=1$ and $\tilde{\alpha}=0.5$.}}
 \label{fig4}
 \end{center}
 \end{figure}
According to Figure \eqref{fig4}, we find that there is only one critical point with a topological charge of +1 $(Q_t=+1)$.\\ 
In the following, for further investigation between these two formalisms, we will draw the $T-S$ or $T-r_h$ diagram and discuss them.

\begin{figure}[h!]
 \begin{center}
 \subfigure[]{
 \includegraphics[height=6cm,width=6cm]{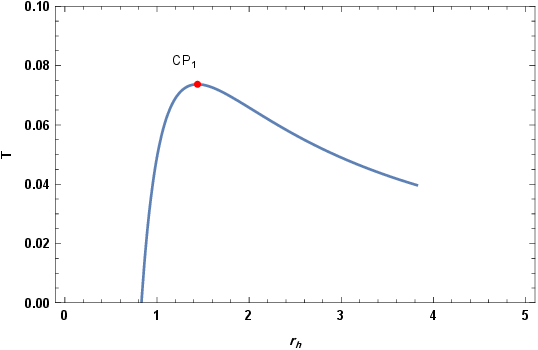}
 \label{fig5a}}
 \subfigure[]{
 \includegraphics[height=6cm,width=6cm]{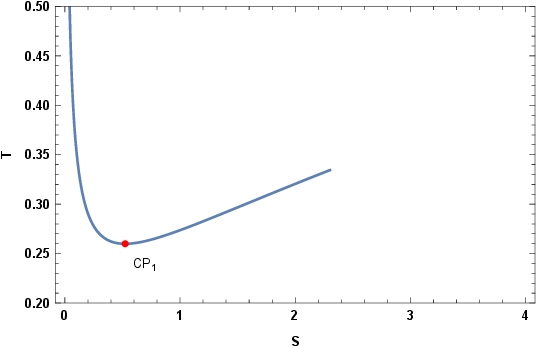}
 \label{fig5b}}
 \caption{\small{(a)Spinodal curve for the R-N AdS black hole in bulk-boundary thermodynamics. With the $Q=C=1$ . (b) Spinodal curve for the R-N AdS black hole in RPS thermodynamics. With the $C=\ell=1$.}}
 \label{fig5}
 \end{center}
 \end{figure}
From Figure \eqref{fig5a}, we can observe that R-N AdS exhibits a maximum point in bulk-boundary thermodynamics with a topological charge of -1. According to proposal \cite{aa}, a first-order phase transition occurs within it. However, for the RPS formalism in Figure \eqref{fig5b}, the critical point is located at the minimum of $T-S$. 
Since the cosmological constant is considered fix in the RPS formalism, we do not associate the black hole with the concept of volume. Therefore, the processes considered in RPS are not related to any volume work. Therefore, in this formalism, there is a behavior similar to that of van der Waals. Also, this critical point can show us the phase transition, which is stated in \cite{fff}.
Such phase transitions are considered as non-equilibrium transitions.
\begin{figure}[h!]
 \begin{center}
 \subfigure[]{
 \includegraphics[height=4cm,width=4cm]{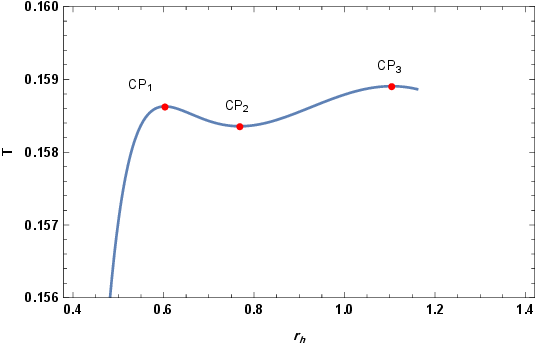}
 \label{fig6a}}
 \subfigure[]{
 \includegraphics[height=4cm,width=4cm]{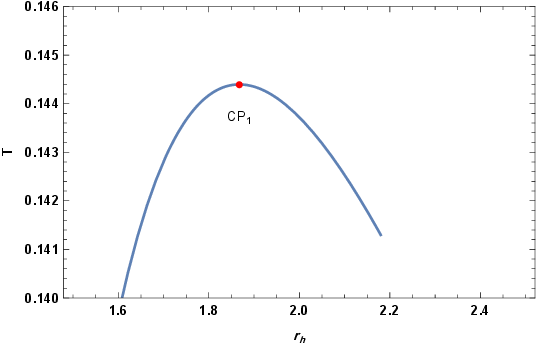}
 \label{fig6b}}
 \subfigure[]{
 \includegraphics[height=4cm,width=4cm]{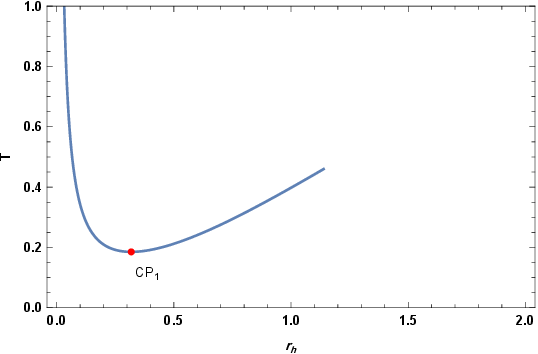}
 \label{fig6c}}
 \caption{\small{Spinodal curve for the Charged Gauss-Bonnet AdS black hole in bulk-boundary thermodynamics. With the  $C=30, \tilde{\alpha}=0.5$,  (a) $Q=0.04$, and (b) $Q=1$ . (c) Spinodal curve for the Charged Gauss-Bonnet AdS black hole in RPS thermodynamics. With the $\ell=1, \tilde{\alpha}=0.5$.}}
 \label{fig6}
 \end{center}
 \end{figure}
From Figures \eqref{fig6a} and \eqref{fig6b}, it is evident that the number of critical points and the total topological charge for charged Gauss-Bonnet AdS black holes in bulk-boundary thermodynamics is identical to that in extended thermodynamics. Furthermore, their phase transitions exhibit similar behavior. However, as shown in Figure \eqref{fig6c}, the behavior of the critical point and its topological charge is entirely distinct between boundary and RPS thermodynamics.\\
In fact, when comparing two black holes in two different thermodynamic systems, we observe that the behavior of the number of critical points and the total topological charge of the black hole is the same in both the boundary and extended thermodynamics. However, in the RPS formalism, their behavior differs between the boundary and extended thermodynamics.
We also find that in the RPS formalism, there is only one critical point with a topological charge of +1 $(Q_t=+1)$.\\
\begin{center}
\begin{table}
  \centering
 \begin{tabular}{|p{5cm}|p{3.5cm}|p{3.2cm}|}
   \hline
   \centering{Case} & \centering{Thermodynamics} &  Total Topological Charge \\
 \hline
  \multirow{3}{5cm}{} & $EPST$  & $Q_{t}=-1$ \\[2mm]
  R-N AdS black hole &$BBT$ & $Q_{t}=-1$ \\[2mm]
  & $RPST$ & $Q_{t}=+1$ \\
   \hline
    \multirow{3}{5cm}{} & $EPST$ & $Q_{t}=-1$   \\[2mm]
     Charged Gauss-Bonnet AdS black hole & $BBT$& $Q_{t}=-1$  \\[2mm]
    &$RPST$ & $Q_{t}=+1$   \\
   \hline
 \end{tabular}
\caption{Summary of the results. Extended phase space thermodynamics (EPST), bulk-boundary thermodynamics (BBT) and restricted phase space thermodynamics (RPST). }\label{20}
\end{table}
 \end{center}
\newpage

\section{Conclusions}
We investigated R-N AdS and charged Gauss-Bonnet AdS black holes in bulk-boundary thermodynamics and RPS thermodynamics. We find that the number of critical points and the total topological charge of the two black holes considered in bulk-boundary thermodynamics are the same as the extended thermodynamics and this result is in agreement with \cite{100}. However, the topological charge of these two black holes in the RPS formalism differs from that of the bulk-boundary and extended thermodynamics. Also, in the RPS formalism, we only find one critical point with a topological charge of +1 for the charged AdS black hole.\\
These three thermodynamics are considered to be the same in one respect: additional terms and parameters to charged AdS black holes do not affect the topological class of the black hole. However, there is a difference between bulk-boundary and extended thermodynamics. In both cases, the total topological charge for the black hole is -1 $(Q_t=-1)$ in all modes.
Conversely, in RPS thermodynamics, the total topological charge is +1 $(Q_t=+1)$.\\
The cosmological constant plays a significant role in the topological charge. Since the cosmological constant is variable in both bulk-boundary and extended thermodynamics, their total topological charges are the same, but in RPS thermodynamics, the cosmological constant is fixed, and hence its total topological charge is different from the other two thermodynamics.

\newpage
\section{Acknowledgments}
We are grateful to Prof Shao-Wen Wei for providing some valuable points.

\end{document}